\documentclass[runningheads]{llncs}
\usepackage[T1]{fontenc}
\usepackage{graphicx}
\usepackage{booktabs}
\usepackage{multirow}
\usepackage{amsmath}
\usepackage{amssymb}
\usepackage{bbding}
\begin{document}
\title{A Lightweight Dual-Branch Visual Front-End for Efficient Audio-Visual Target Speaker Extraction}
%
\titlerunning{Lightweight Dual-Branch Visual Front-End for AVTSE}
%

\author{Zhan Jin\inst{1,2} \and
Peijun Yang\inst{3,4} \and
Juan Liu\inst{3,4} \and
Ming Li\inst{3,2}\thanks{Corresponding author: Ming Li}}
\authorrunning{Z. Jin et al.}

\institute{School of Computer Science, Wuhan University, Wuhan, China \and
School of Artificial Intelligence, Chinese University of Hong Kong, Shenzhen, China \and
School of Artificial Intelligence, Wuhan University, Wuhan, China \and
School of Cyber Science and Engineering, Wuhan University, Wuhan, China\\\email{mingli369@cuhk.edu.cn}}

\maketitle              

\begin{abstract}
Audio-Visual Target Speaker Extraction (AVTSE) uses face-related visual cues to extract a target speaker from a speech mixture. In many AVTSE systems, visual features are obtained from convolutional image encoders, and their computation may become a non-negligible part of the overall system when lightweight separation backbones are used. This work proposes a dual-branch visual front-end that reorganizes visual information according to its spatial characteristics. A compact lip region of interest is used to model local articulation cues through convolutional processing, while facial landmark coordinates are used to represent global geometric motion with a lightweight temporal convolutional network. The two representations are integrated into IIANet and SpatialNet with limited fusion overhead. Experiments on LRS3-2mix under a unified reduced training protocol show that the proposed method achieves a performance–computation trade-off close to the lightweight image-based baseline. Compared with the ResNet18-based visual encoder, it reduces the module-level visual encoding computation by more than 90\% across both evaluated separation backbones, with a moderate performance gap. This comparison excludes face detection and landmark estimation and therefore applies to systems where landmark coordinates are already available from preprocessing. Demos are available at \url{https://anonymous.4open.science/r/Lightweight-Visual-Frontend-E46E}.

\end{abstract}

\section{Introduction}

Audio-Visual Target Speaker Extraction (AVTSE) is a task that uses face-related visual cues to assist deep learning networks in extracting the target speaker's speech from a speech mixture. This task is inspired by human multimodal perception, where visual information such as lip movements and facial motions often helps improve speech understanding in complex acoustic environments. With the growing demand for audio-visual perception systems in scenarios such as complex interaction, intelligent meetings, hearing-assistive devices, and robotic audition, AVTSE has received increasing attention in recent years and has achieved continuous progress in model architecture and separation performance.

Wu et al. introduced a time-domain architecture into audio-visual speech separation and proposed an end-to-end time-domain audio-visual separation network. In two-speaker and three-speaker mixture scenarios, their method reported higher signal-to-noise ratio improvement than an audio-driven Conv-TasNet baseline \cite{avtasnet}. Gao and Grauman proposed the VisualVoice framework, which adopts a multi-task learning strategy. While learning audio-visual speech separation from unlabeled videos, it jointly learns cross-modal speaker embeddings from lip motion and facial appearance, thereby improving generalization to real-world video scenarios \cite{visualvoice}. Inspired by the neuroanatomical structure of mammalian cortico-thalamo-cortical circuits, Li et al. proposed CTCNet \cite{ctcnet} which performs multi-round audio-visual fusion through hierarchical representation learning and top-down connections in the thalamic subnetworks, achieving better separation results than several existing methods on multiple benchmark datasets with fewer parameters. Pan et al. proposed SAV-TFGridNet \cite{savgridnet} based on the time-frequency-domain separation model TF-GridNet \cite{tfgridnet}. Their method incorporates the target speaker's facial recording as conditional information into the separation process and introduces an interference-type identification mechanism to adapt expert models to different scenarios, achieving competitive results in the COG-MHEAR audio-visual speech enhancement challenge \cite{avsec}. AV-MossFormer2, as one of the core components of the ClearerVoice-Studio open-source toolkit, implements end-to-end multimodal speech processing for target speaker extraction based on an improved MossFormer2 architecture \cite{avmossformer2}. However, from the perspective of practical deployment, performance improvements are often accompanied by increased computational and storage costs, which still limits the application of AVTSE models on low-computing-power devices.

In recent years, several studies have begun to explore network architectures with lower computational complexity. Pegg et al. proposed RTFS-Net \cite{rtfsnet}, which operates on complex time-frequency bins obtained from the short-time Fourier transform. It models the time and frequency dimensions separately using multilayer recurrent neural networks and introduces an attention-based audio-visual fusion mechanism. According to the original report, RTFS-Net achieves better separation results with a smaller fraction of the parameters and multiply-accumulate operations of the baseline model. IIANet \cite{iianet} further employs intra-modal attention and cross-modal attention for efficient audio-visual feature fusion. Its fast version achieves favorable separation quality with lower multiply-accumulate operations and faster CPU inference. These methods mainly reduce the overall computational complexity from the perspective of the separation backbone. However, in most AVTSE systems, the visual front-end still accounts for a considerable proportion of image feature extraction computation. As separation backbones become increasingly lightweight, the proportion of the visual front-end in the total computation may further increase, making it an important limiting factor for low-computing-power deployment.

To address the computational cost of visual front-ends, existing studies have started to design visual front-ends with lower computational loads. Martel et al. proposed AVLIT, which adopts a progressive learning strategy and an asynchronous fully recurrent convolutional neural network. With a lightweight iterative architecture, it achieves better results than several audio-only and audio-visual baselines in noisy audio-visual speech separation tasks \cite{avlit}. Li et al. proposed Dolphin, which designs DP-LipCoder, a dual-path lightweight video encoder for visual feature extraction. It converts lip motion into audio-aligned discrete semantic tokens and embeds global-local attention modules in the separation network to capture multi-scale dependencies. According to the original report, Dolphin improves separation quality while reducing the number of parameters and achieving notable GPU inference acceleration \cite{dolphin}. Pan et al. proposed an online audio-visual autoregressive speaker extraction model, which combines a lightweight visual front-end with an autoregressive acoustic encoder in streaming scenarios and achieves performance close to relevant baselines with reduced computational scale \cite{blazenet}. These studies indicate that reducing the computational cost of the visual front-end is practically meaningful for AVTSE systems. Nevertheless, as lower-computation separation backbones continue to emerge, the computational load of the visual front-end still needs to be further controlled.

In this work, we use attention heatmap visualization as a preliminary observation tool to inspect how an AVTSE model distributes attention over different locations in the facial region of interest. The visual examples suggest that the inner lip region tends to receive concentrated responses, whereas several peripheral facial regions, including the outer lip, subnasal area, and jaw, may provide more spatially sparse auxiliary cues. Based on this observation, we consider a trade-off between two visual modeling strategies. Modeling a large facial region with two-dimensional convolutions introduces computation over areas that may be weakly related to the extraction task, whereas modeling only a compact lip region may remove auxiliary facial motion. Considering that facial landmarks are commonly used for face alignment and mouth localization, and that they provide a low-dimensional geometric representation, we propose a dual-branch visual front-end. The proposed front-end uses a small lip ROI to model local appearance and articulation cues, and uses landmark sequences to model global facial dynamics. The objective is not to improve the upper bound of separation performance, but to reduce the additional computation of the visual front-end under a controlled comparison with existing visual front-ends.

\section{Related Work}

Visual front-ends based on pretrained lip-reading models are widely adopted in AVTSE. A typical architecture consists of a 3D convolutional layer followed by ResNet18, whose weights are usually pretrained on lip-reading datasets before being integrated into speech separation networks \cite{lipreading}. Although this architecture provides effective visual representations, its computational cost is non-negligible. As separation backbones become increasingly lightweight, the computation introduced by the visual front-end accounts for a growing proportion of the overall system, limiting deployment on resource-constrained platforms.

To reduce the computational burden of visual feature extraction, several lightweight visual front-ends have been proposed. BlazeNet64 adopts the efficient convolutional design of BlazeFace \cite{blazeface} to reduce the computation required for visual encoding while maintaining competitive separation performance \cite{blazenet}. Dolphin further introduces the lightweight DP-LipCoder, which represents lip motion using compact discrete semantic tokens to reduce the dependence on high-dimensional visual features \cite{dolphin}. Although these methods substantially improve the efficiency of image-based visual encoding, the visual front-end still contributes a considerable fraction of the total computation in lightweight AVTSE systems.

Besides designing lightweight image encoders, alternative visual representations have also been explored. PIAVE incorporates an additional facial representation generated by SADRNet into the separation framework \cite{piave}. During this process, SADRNet estimates facial landmarks as intermediate outputs for face alignment and facial reconstruction \cite{sadrnet}. While PIAVE mainly utilizes the synthesized facial view, the landmark coordinates provide a compact geometric description of facial motion. Different from methods that rely on a single image-based visual representation, this work explicitly separates local lip appearance modeling from global geometric motion modeling. The proposed design is therefore positioned as a representation allocation strategy for reducing the computation of the visual front-end inside AVTSE systems.

    
    

\begin{figure}[t]
    \centering
    \begin{tabular}{@{}cc@{}}
        \includegraphics[width=0.28\linewidth]{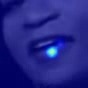} &
        \includegraphics[width=0.28\linewidth]{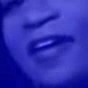} \\
        (a) Head 1 & (b) Head 2 \\
        \includegraphics[width=0.28\linewidth]{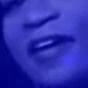} &
        \includegraphics[width=0.28\linewidth]{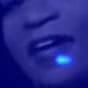} \\
        (c) Head 3 & (d) Head 4
    \end{tabular}
    \caption{Spatial attention maps from four attention heads.
    The $11\times11$ feature maps are bilinearly upsampled to the original
    $88\times88$ lip ROI for visualization.}
    \label{fig:grid}
\end{figure}

\section{Methods}

\subsection{Preliminary Observation and Motivation}

To examine the spatial distribution of visual responses in AVTSE models, a visual self-attention module is inserted between the ResNet18 visual front-end and the audio-visual fusion module. Different from the original setting, where global average pooling is applied before feature fusion, the intermediate spatial feature map is preserved for visualization. For an $88\times88$ face ROI, ResNet18 produces a feature map with spatial resolution $11\times11$. After flattening the spatial dimensions, the multi-head self-attention module computes
\begin{equation}
\mathbf{A}
=
\operatorname{Softmax}
\left(
\frac{\mathbf{Q}\mathbf{K}^{\top}}
{\sqrt{d}}
\right),
\end{equation}
where $\mathbf{A}\in\mathbb{R}^{121\times121}$ denotes the attention relationship among spatial locations. For visualization, the attention responses are averaged over all query positions, reshaped into an $11\times11$ map, and upsampled to the original ROI resolution by bilinear interpolation.

Figure~\ref{fig:grid} suggests heterogeneous spatial response patterns: one response is concentrated around the inner lip region, while others are more spatially dispersed around the outer lip, subnasal region, and jaw. This observation motivates a representation-allocation hypothesis: dense local articulation cues may benefit from image-based modeling, whereas spatially distributed facial motion may be represented more compactly by geometric coordinates. The visualization is used only as qualitative motivation and does not establish causal importance of individual regions.

Based on this hypothesis, we use a compact lip ROI to preserve local appearance information and facial landmarks to describe broader geometric motion.

\begin{figure}[t]
    \centering
    \includegraphics[width=\linewidth]{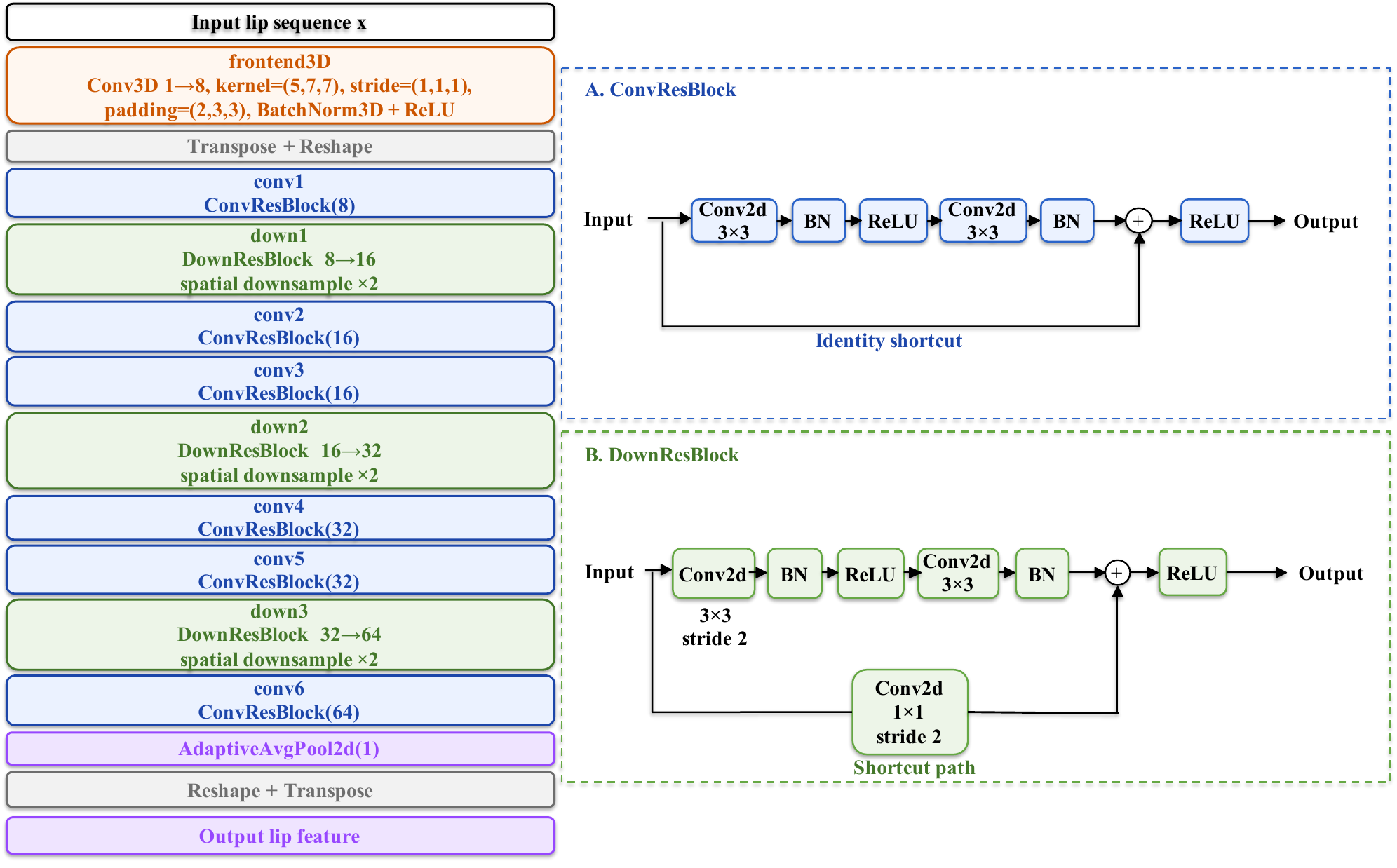}
    \caption{Architecture of the Lip Encoder}
    \label{fig:lipencoder}
\end{figure}

\subsection{Lip Motion Modeling Branch}

In the lip motion modeling branch, a local lip image patch is cropped from the original $96\times96$ facial ROI. The mean coordinate of the lip contour landmarks is used as the crop center, and a local window of $36\times36$ pixels is first extracted. During training, a $32\times32$ lip ROI is randomly cropped from this local window, and horizontal flipping is applied as data augmentation. During validation and testing, center cropping is used. This compact ROI is adopted to retain the main local lip motion while reducing computation over peripheral facial regions. In the ablation study, a horizontally extended $32\times64$ ROI is also evaluated. Here, $32\times64$ denotes height by width. This setting is included because a $32\times32$ crop may miss horizontally extended mouth shapes, whereas the wider crop can cover the complete lip contour along the horizontal direction.

The cropped lip image patch is processed by the Lip Encoder shown in Figure~\ref{fig:lipencoder}. The implementation contains an initial three-dimensional convolutional layer for short-term spatio-temporal modeling, followed by batch normalization and nonlinear activation. The resulting features are reshaped into frame-level feature maps and passed through a ResNet-style two-dimensional convolutional stack composed of spatial downsampling residual blocks and channel-preserving residual blocks. After spatial aggregation, a temporal projection layer produces a 64-channel visual feature sequence. In compact form, the structure is
\begin{equation}
\begin{aligned}
\mathrm{Conv3D}
&\rightarrow \mathrm{BN}
\rightarrow \mathrm{ReLU}\\
&\rightarrow \mathrm{2D\ ResNet\text{-}style\ blocks}\\
&\rightarrow \mathrm{Spatial\ Pooling}
\rightarrow \mathrm{Temporal\ Projection}.
\end{aligned}
\end{equation}
The module is trained end to end for target speaker extraction. Compared with conventional $88\times88$ ROI-based encoders, this branch reduces convolutional computation mainly by limiting the input spatial range rather than by changing the basic convolutional modeling principle.

\begin{figure}[t]
    \centering
    \includegraphics[width=\linewidth]{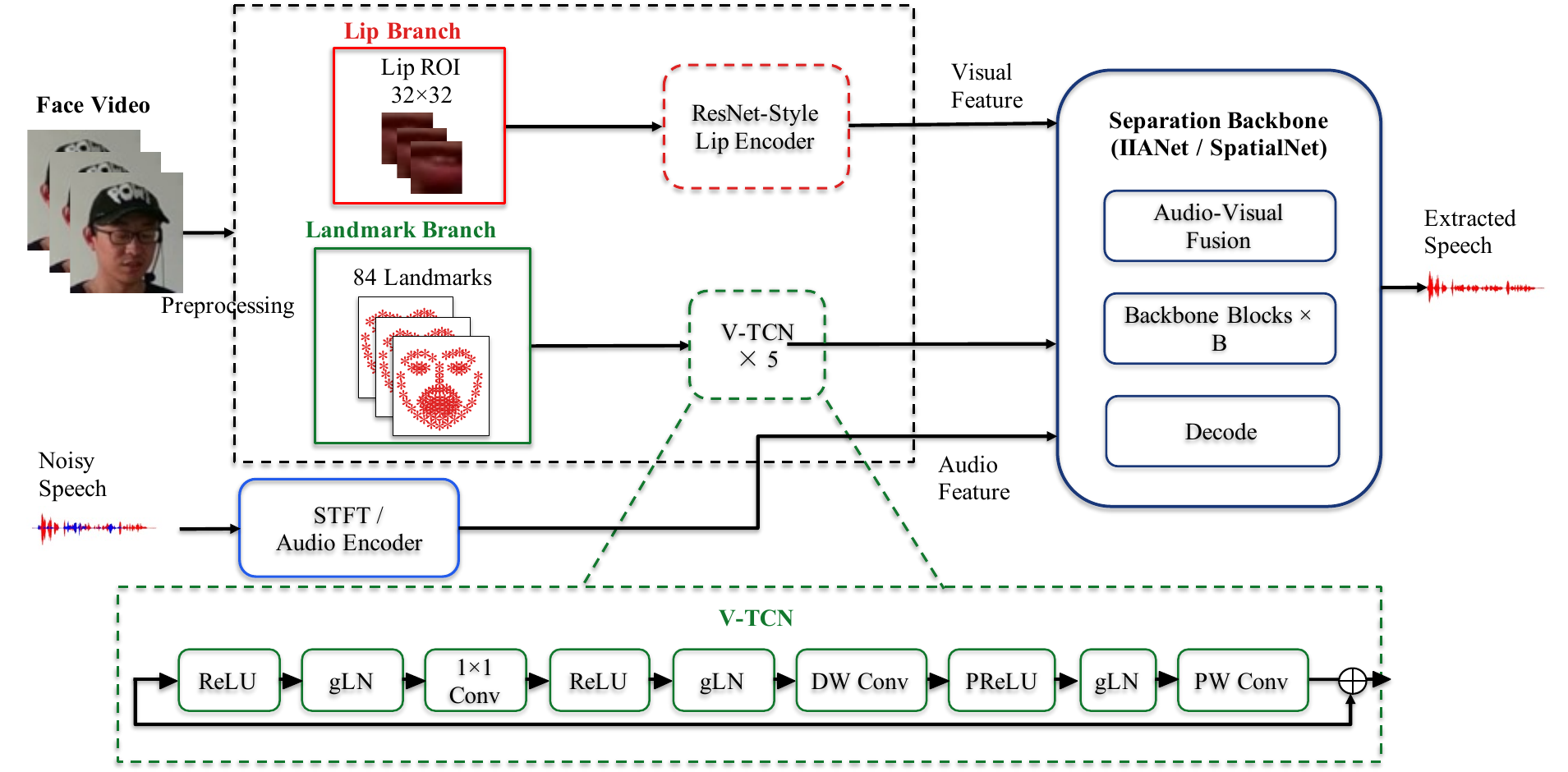}
    \caption{Overall architecture of the proposed method}
    \label{fig:framework}
\end{figure}

\subsection{Facial Landmark Modeling Branch}

Compared with image-based visual representations, facial landmark coordinates provide a compact description of facial geometry and motion. Facial landmarks are commonly produced during face alignment and mouth ROI localization and can therefore be reused when such preprocessing is already available. We adopt SADRNet \cite{sadrnet}, as in PIAVE \cite{piave}, because landmark coordinates are directly used as network input. Under the same THOP protocol, SADRNet and FAN-2D \cite{fan2d} require 10.4 and 25.4 GMACs per 256 × 256 image, respectively. Landmark estimation is treated as preprocessing and is excluded from the visual encoding costs reported in Tables~\ref{tab:iianet_main}--\ref{tab:ablation}; if it must be newly introduced, its cost should be included in end-to-end deployment analysis.

The standard 68-point template already describes the main facial contour and lip boundary, but provides relatively sparse sampling in the peripheral regions highlighted by the qualitative visualization. We therefore add 16 points around the upper-lip/subnasal and lip-to-jaw regions to increase geometric sampling density in these areas, resulting in an 84-point representation. Specifically, 8 points are placed between the nose and upper lip, and 8 between the mouth corners and jaw contour. The additional landmark locations were selected to provide denser geometric coverage of the peripheral regions suggested by the attention visualization, rather than to imply that these regions are causally more important. The two-dimensional coordinates of all landmarks are concatenated frame by frame and processed by V-TCN blocks \cite{usev} to model temporal facial dynamics.

Let $\mathbf{p}_{n,t,k}\in\mathbb{R}^{2}$ denote the coordinate of the $k$-th landmark in the $t$-th frame of the $n$-th training sample. Landmark normalization is performed independently for each landmark index and each coordinate dimension. The mean and standard deviation are computed over all frames in the training set as
\begin{equation}
\mu_{k,c}=\frac{1}{N_{\mathrm{tr}}}\sum_{n,t} p_{n,t,k,c},
\end{equation}
\begin{equation}
\sigma_{k,c}=\sqrt{\frac{1}{N_{\mathrm{tr}}}\sum_{n,t}\left(p_{n,t,k,c}-\mu_{k,c}\right)^2+\epsilon},
\end{equation}
where $c\in\{x,y\}$ and $N_{\mathrm{tr}}$ denotes the total number of frames in the training set. The normalized coordinate is then computed as
\begin{equation}
\tilde{p}_{n,t,k,c}=\frac{p_{n,t,k,c}-\mu_{k,c}}{\sigma_{k,c}}.
\end{equation}
The statistics $\mu_{k,c}$ and $\sigma_{k,c}$ are computed only from the training set and are reused for validation and testing.

During data augmentation, horizontal flipping is applied consistently to both the lip image and the corresponding landmark coordinates. The landmark transformation is performed before normalization. Specifically, the horizontal coordinate is mirrored according to the ROI width, and the left--right symmetric landmark indices are exchanged accordingly. This permutation is applied to both the original 68 facial landmarks and the additional 16 landmarks, so that each landmark channel preserves a consistent semantic meaning after augmentation.

\subsection{Integration with Separation Backbones}

Figure~\ref{fig:framework} shows the integration of the proposed dual-path visual front-end with IIANet and SpatialNet. IIANet performs hierarchical audio-visual interaction through intra-modal and cross-modal attention mechanisms \cite{iianet}. In this backbone, the proposed visual front-end replaces the original visual front-end while the remaining network architecture and fusion strategy are kept unchanged. SpatialNet is a time-frequency-domain separation network composed of alternating narrow-band and cross-band modules for modeling temporal, spectral, and spatial correlations \cite{spatialnet}. Following the implementation in \cite{avspatialnet}, the proposed visual front-end is incorporated as conditional visual information without modifying the original backbone architecture. These two backbones are used to examine whether the proposed visual representation can be integrated into different separation frameworks under a unified evaluation setting.

\subsection{Visual Feature Fusion}

The proposed visual front-end produces complementary features from the lip motion branch and the landmark branch. For IIANet, both branches produce 64-channel visual features. The two features are concatenated along the channel dimension and projected back to the original 64-dimensional feature space using a single $1$D convolution before being fed into the original audio-visual fusion module. For SpatialNet, the visual feature dimensions are aligned with the frequency dimension of the input spectrogram. The lip feature is projected from 64 to 129 channels, while the V-TCN is configured to output 129-channel landmark features. The two visual features are then concatenated with the complex spectrogram along the introduced visual dimension, enabling the backbone to use both visual representations without changing its main separation architecture.

\section{Experiments}\label{sec:experiments}

\subsection{Datasets}

The experiments are conducted on the public LRS3-2mix dataset, following the CTCNet-related setting \cite{ctcnet} to construct the training, validation, and test sets. The dataset contains 50,000 training mixtures, 5,000 validation mixtures, and 3,000 evaluation mixtures. Each mixture is generated by mixing two clean speech signals of 2 seconds at a sampling rate of 16 kHz with a random signal-to-noise ratio ranging from -5 dB to 5 dB. The corresponding original video is a face video with a frame rate of 25 FPS and a spatial size of $224\times224$. After preprocessing, a mouth patch window of $96\times96$ is obtained and further used to construct the inputs for different visual front-ends.

\subsection{Training Configuration}

The proposed visual front-end is compared with two representative image-based visual front-ends, namely the conventional pretrained ResNet18 encoder \cite{lipreading} and the lightweight BlazeNet64 encoder \cite{blazenet}. Both baselines use an $88\times88$ mouth ROI as input. During training, the input is randomly cropped from a $96\times96$ mouth patch with horizontal flipping applied for data augmentation. During validation and testing, center cropping is adopted.

For the proposed dual-branch visual front-end, the lip branch first extracts a $36\times36$ local region centered on the mouth area, which is further randomly cropped to $32\times32$ during training. The same horizontal flipping operation is applied consistently to both the lip image branch and the landmark branch. As described in Section~3.3, landmark channels are also permuted after flipping to preserve left--right semantic consistency. All mixture speech signals are pre-generated offline, and no dynamic mixing is used.

The training objective follows a scale-invariant waveform-level loss with an auxiliary scale-adjusted $L_1$ term. Given the target signal $\mathbf{s}\in\mathbb{R}^{T}$ and the estimated signal $\hat{\mathbf{s}}\in\mathbb{R}^{T}$, the scale factor is defined as
\begin{equation}
\alpha=\frac{\hat{\mathbf{s}}^{\mathsf T}\mathbf{s}}{\hat{\mathbf{s}}^{\mathsf T}\hat{\mathbf{s}}},
\end{equation}
and the loss is computed as
\begin{equation}
\mathcal{L}_{\mathrm{SI\text{-}SDR}}
=
-10\log_{10}
\frac{\|\mathbf{s}\|_2^2}
{\|\alpha\hat{\mathbf{s}}-\mathbf{s}\|_2^2},
\end{equation}
\begin{equation}
\mathcal{L}
=
\mathcal{L}_{\mathrm{SI\text{-}SDR}}
+
\lambda
\left\|
\alpha\hat{\mathbf{s}}-\mathbf{s}
\right\|_{1},
\end{equation}
where $T$ denotes the number of samples. The weighting factor $\lambda$ is set to 1 in all experiments.

All models are trained on 8 V100 GPUs with a batch size of 3. The optimizer is Adam with an initial learning rate of 0.001. The learning rate is halved when the validation metric does not improve within a predefined tolerance, and training is stopped when the early stopping criterion is reached.

Due to computational constraints, a reduced training protocol is adopted instead of strictly following the original IIANet setting. Specifically, one of the two clean speech signals is randomly selected as the training target for each mixture, and two training schedules with different tolerance/early-stopping configurations are used (3/10 and 5/15). All methods in the same comparison group are trained under identical conditions. Therefore, the reported results are intended for relative comparison among visual front-ends under a unified setting, rather than for refreshing benchmark performance on LRS3-2mix.

\subsection{Evaluation Metrics}

The evaluation metrics include SI-SNR improvement, PESQ, and \ multiply-accumulate operations. Let $\mathbf{y}$ denote the mixture signal, $\mathbf{s}$ denote the clean target signal, and $\hat{\mathbf{s}}$ denote the estimated target signal. Following common speech separation evaluation practice \cite{sisdr}, SI-SNR improvement is computed as
\begin{equation}
\mathrm{SI\text{-}SNRi}(\mathbf{y},\mathbf{s},\hat{\mathbf{s}})
=
\mathrm{SI\text{-}SNR}(\mathbf{s},\hat{\mathbf{s}})
-
\mathrm{SI\text{-}SNR}(\mathbf{s},\mathbf{y}).
\end{equation}
For two signals $\mathbf{u}$ and $\mathbf{v}$, where $\mathbf{u}$ denotes the reference signal and $\mathbf{v}$ denotes the evaluated signal, SI-SNR is defined as
\begin{equation}
\mathrm{SI\text{-}SNR}(\mathbf{u},\mathbf{v})
=
10\log_{10}\frac{\|\mathbf{v}_{\mathrm{target}}\|_2^2}
{\|\mathbf{e}_{\mathrm{noise}}\|_2^2},
\end{equation}
where
\begin{equation}
\mathbf{v}_{\mathrm{target}}
=
\frac{\langle \mathbf{v},\mathbf{u}\rangle}{\|\mathbf{u}\|_2^2}\mathbf{u},
\qquad
\mathbf{e}_{\mathrm{noise}}
=
\mathbf{v}-\mathbf{v}_{\mathrm{target}}.
\end{equation}
PESQ is used as a perceptual quality metric \cite{pesq}. Computational complexity is measured using the \texttt{profile} function of the THOP library and reported in GMACs. The reported values measure module-level visual encoding computation, excluding video decoding, face detection \cite{retinaface,mtcnn}, ROI extraction, and landmark estimation \cite{sadrnet}.

\begin{table}[t]
\centering
\caption{Performance comparison of different visual front-ends on the LRS3-2mix test set using the IIANet backbone. GMACs are reported as backbone + visual encoding. The base IIANet backbone requires 17.1 GMACs.}
\label{tab:iianet_main}
\begin{tabular}{ccp{3.7cm}ccc}
\toprule
Tolerance & Early stop & Visual front-end & SI-SNRi & PESQ & GMACs \\
\midrule
\multirow{3}{*}{5} & \multirow{3}{*}{15}
& ResNet18(pretrained) \cite{lipreading} & 17.5 & 2.90 & 17.1+8.0 \\
& & BlazeNet64~\cite{blazenet} & 17.0 & 2.80 & 17.1+2.4 \\
& & 32$\times$32 + lmk84 & 17.1 & 2.80 & 17.1+0.3 \\
\midrule
\multirow{3}{*}{3} & \multirow{3}{*}{10}
& ResNet18(pretrained) & 16.2 & 2.64 & 17.1+8.0 \\
& & BlazeNet64 & 16.0 & 2.58 & 17.1+2.4 \\
& & 32$\times$32 + lmk84 & 16.0 & 2.59 & 17.1+0.3 \\
\bottomrule
\end{tabular}
\end{table}

\begin{table}[t]
\centering
\caption{Performance comparison of different visual front-ends on the LRS3-2mix test set using the SpatialNet backbone. All models are trained with $tolerance=3$ and $early\_stop=10$. GMACs are reported as backbone + visual encoding. The base SpatialNet backbone requires 17.4 GMACs.}
\label{tab:spatialnet}
\begin{tabular}{p{3.5cm}ccc}
\toprule
Visual front-end & SI-SNRi & PESQ & GMACs \\
\midrule
ResNet18 (pretrained) & 15.6 & 2.79 & 17.4+8.0 \\
BlazeNet64 & 15.5 & 2.77 & 17.4+2.9 \\
32$\times$32 + lmk84 & 15.5 & 2.81 & 17.4+0.5 \\
\bottomrule
\end{tabular}
\end{table}

\begin{table}[t]
\centering
\caption{Ablation study of the proposed visual front-end on the LRS3-2mix test set using the IIANet backbone. All models are trained with $tolerance=3$ and $early\_stop=10$. GMACs are reported as backbone + visual encoding. The base IIANet backbone requires 17.1 GMACs.}
\label{tab:ablation}
\begin{tabular}{p{3.5cm}ccc}
\toprule
Configuration & SI-SNRi & PESQ & GMACs \\
\midrule
32$\times$32 ROI + lmk84 & 16.0 & 2.59 & 17.1+0.3 \\
32$\times$32 ROI only & 14.9 & 2.38 & 17.1+0.3 \\
32$\times$64 ROI only & 15.0 & 2.32 & 17.1+0.6 \\
lmk84 only & 15.3 & 2.59 & 17.1+0.01 \\
lmk68 only & 15.2 & 2.58 & 17.1+0.01 \\
\bottomrule
\end{tabular}
\end{table}

\subsection{Experiment Results}

Table~\ref{tab:iianet_main} presents the experimental results of different visual front-ends using the IIANet backbone. Under the unified reduced training protocol, the proposed $32\times32+lmk84$ dual-branch visual front-end reduces the module-level computation of the visual front-end while maintaining a separation performance close to BlazeNet64.

Under the $tolerance=5$ and $early\_stop=15$ setting, the proposed method achieves the same PESQ as BlazeNet64 and a slightly higher SI-SNRi value. Since no multi-seed evaluation is conducted, this small numerical difference should not be interpreted as a statistically verified improvement. The main observation is that the proposed method remains close to BlazeNet64 in separation performance while reducing the module-level visual encoding cost from 2.4 to 0.3 GMACs under the same THOP protocol.

Compared with the pretrained ResNet18 visual front-end, the proposed method shows a performance gap. This result suggests that pretrained image-based encoders still retain stronger capability in modeling fine-grained visual appearance. Nevertheless, the proposed method reduces the module-level visual front-end computation from 8.0 to 0.3 GMACs under the IIANet setting. The result therefore reflects a trade-off between computation and separation quality, rather than an improvement in the upper bound of performance.

Under the more restricted $tolerance=3$ and early stopping $=10$ setting, all visual front-ends show lower scores. The proposed method remains numerically close to BlazeNet64, but the small differences among lightweight encoders should be interpreted as empirical trends under the current setup. Since the experiments adopt fixed mixture pairs, random target selection, and a reduced training schedule, the results are not intended to reproduce the original best results of IIANet.

Table~\ref{tab:spatialnet} reports results on the SpatialNet backbone. A similar tendency is observed: the proposed method requires less module-level visual encoding computation than the two image-based front-ends, while achieving comparable SI-SNRi and PESQ under the current setting. Together with the IIANet results, this shows that the proposed representation can be integrated into two different separation backbones, although broader architectural generality is not established by the current evaluation.

Overall, the experimental results support the relative comparison objective of this work. The results demonstrate the intended performance--computation trade-off in module-level visual encoding under the evaluated setting.

\subsection{Ablation Study}

Table~\ref{tab:ablation} presents the ablation results of the proposed dual-branch visual front-end. Compared with using only the $32\times32$ ROI, the full configuration combining the $32\times32$ ROI and lmk84 improves SI-SNRi from 14.9 dB to 16.0 dB and PESQ from 2.38 to 2.59. This indicates that, when the lip image region is strongly compressed, landmark coordinates provide complementary geometric information at a low computational cost.

The $32\times64$ ROI-only configuration tests whether the 32 × 32 model is mainly limited by horizontal mouth coverage. Although the wider ROI slightly improves SI-SNRi from 14.9 to 15.0 dB, it remains below the dual-branch model and doubles the image-branch computation. This suggests that the gain of the proposed method cannot be explained simply by wider mouth coverage.

When only the lmk68 branch is used, the model achieves 15.2 dB SI-SNRi and 2.58 PESQ. The lmk84-only model achieves 15.3 dB SI-SNRi and 2.59 PESQ, outperforming the 32 × 32 ROI-only model (14.9 dB / 2.38). This result indicates that, under the current compact-ROI setting, landmark geometry alone provides competitive target-speaker cues relative to the compact image branch. Adding the lip-image branch further improves SI-SNRi to 16.0 dB, while PESQ remains at 2.59. The additional gain suggests that the lip-image branch provides information not fully captured by landmark geometry, potentially including fine-grained articulation and mouth-interior appearance. Therefore, the two branches provide complementary representations under the evaluated setting.

The ablation results are consistent with the design hypothesis that geometric facial motion can provide a compact complement to dense local image features. Since the ablation is conducted under a limited training protocol and without multi-seed evaluation, the conclusion is restricted to empirical trends observed in the current setting.

\section{Discussion}

The experimental results suggest that, under the evaluated setting, visual encoding computation can be reduced by reorganizing visual representations rather than relying solely on more compact image convolutional networks. A ResNet18-based visual front-end can model a relatively large mouth ROI and achieves stronger separation performance in the reported experiments. However, its module-level computation is non-negligible when it is paired with lightweight separation backbones. The proposed method instead decomposes visual information into local appearance cues and global geometric cues. The inner lip region is modeled using a compact ROI, while broader facial motion is represented by landmark coordinates. This design reduces the amount of dense image convolution required by the visual encoder.

Compared with lightweight image-based encoders such as BlazeNet64, the proposed method differs mainly in the representation strategy. The inner lip region contains fine-grained texture and articulation patterns and is therefore modeled by a small image branch. Peripheral regions such as the outer lip, subnasal area, and jaw may provide auxiliary motion information, but processing them through a larger image ROI also increases convolutional computation. Landmark coordinates offer a compact alternative for representing such geometric variations. The ablation study supports this interpretation under the evaluated setting, since the combination of $32\times32$ ROI and lmk84 outperforms either branch alone.

The experiments on IIANet and SpatialNet indicate that the proposed visual representation can be inserted into two different audio-visual separation backbones. IIANet uses cross-modal attention for audio-visual interaction, whereas SpatialNet relies on time-frequency-domain modeling with visual conditioning. The similar trend observed on these two backbones suggests architectural compatibility, but it does not by itself establish broad generality across AVTSE systems. Further evaluation on additional architectures and datasets would be required for a stronger conclusion.

The reported GMACs quantify module-level visual encoding rather than the complete video-processing pipeline; if landmark estimation must be newly introduced, its cost should be included in end-to-end analysis. The landmark branch may also depend on localization quality, while landmark noise, occlusion, large pose variation, and alignment failures are not explicitly evaluated. Therefore, the reported efficiency and robustness should be interpreted within these evaluation boundaries, and THOP-based GMACs should not be regarded as hardware-specific latency.

Overall, the proposed method should be viewed as a trade-off design for low-computation AVTSE visual encoding. Its main contribution lies in combining compact local image modeling with low-dimensional geometric modeling. The current results support this direction under a unified LRS3-2mix evaluation setting, while broader robustness and end-to-end efficiency require further study.

\section{Conclusion}

This paper presents a dual-branch visual front-end for efficient Audio-Visual Target Speaker Extraction. The proposed design combines a compact lip ROI branch for local appearance modeling with a landmark branch for low-dimensional facial motion modeling. Experiments on LRS3-2mix under a unified reduced training protocol show that the proposed front-end achieves a performance--computation trade-off close to the evaluated lightweight image-based visual front-end while reducing module-level visual encoding computation. The ablation results suggest that lip appearance and landmark-based geometric motion provide complementary cues. The current evaluation is limited to controlled settings and does not cover multi-seed training, cross-dataset testing, real-world scenarios, modality-missing conditions, or full end-to-end latency including landmark detection. These aspects remain to be further investigated to better assess the robustness, generalizability, and deployment efficiency of the proposed representation.

\subsubsection{Acknowledgements}
This research is funded in part by the National Natural Science Foundation of China (62571223) and Yangtze River Delta Science and Technology Innovation Community Joint Research Project (2024CSJGG01100).

\subsubsection{Disclosure of Interests}
The authors have no competing interests to declare that are relevant to the content of this article.

%
%
%
%

\end{document}